\documentclass[11pt,a4paper]{article}

\usepackage{amssymb,amsmath,amsfonts}
\usepackage[dvipsnames]{xcolor}
\usepackage{graphicx}
\usepackage{longtable}
\usepackage{verbatim}
\usepackage{color}
\usepackage[mathscr]{euscript}
\usepackage{framed}
\usepackage{mdframed}
\usepackage{amsmath}
\usepackage{amsfonts}
\usepackage{mathrsfs}
\usepackage{amssymb}
\usepackage{mathrsfs}  
\usepackage{subcaption}
\usepackage{caption}
\usepackage{tcolorbox}
\usepackage{mathtools}

\usepackage[left=26mm,top=20mm,right=26mm,bottom=20mm]{geometry}

\usepackage{marvosym} % For cool symbols.
\usepackage{cite}
\usepackage{hyperref}
\hypersetup{
     colorlinks = true,
     colorlinks=true,
     linkcolor = red,
     anchorcolor = blue,
     citecolor = blue,
     filecolor = blue,
     urlcolor = blue
     }
     
\usepackage{tcolorbox}
\usepackage{color}
\usepackage{cite}
\usepackage{slashed}
\usepackage{hyperref}
\hypersetup{colorlinks, citecolor=bluscuro, linkcolor=bluscuro, urlcolor=bluscuro}
\definecolor{rossos}{cmyk}{0,1,1,0.55}
\definecolor{bluscuro}{rgb}{0.15, 0.2, .85}
\definecolor{bluchiaro}{cmyk}{1,.3,0.,0.1}

\graphicspath{{./Figures/}}

\newcommand{\be}{\begin{equation}}
\newcommand{\ee}{\end{equation}}
\newcommand{\bea}{\begin{eqnarray}}
\newcommand{\eea}{\end{eqnarray}}
\newcommand{\beq}{\begin{equation}}
\newcommand{\eeq}{\end{equation}}

\def\beqa{\begin{eqnarray}}

\def\eeqa{\end{eqnarray}}

\def\lsim{\mathrel{\rlap{\lower4pt\hbox{\hskip0.5pt$\sim$}}
    \raise1pt\hbox{$<$}}}         %less than or approx. symbol
\def\gsim{\mathrel{\rlap{\lower4pt\hbox{\hskip0.5pt$\sim$}}
    \raise1pt\hbox{$>$}}}         %greater than or approx. symbol

\usepackage[normalem]{ulem}

\usepackage[greek,english]{babel}
\usepackage[makeroom]{cancel}

\usepackage{pifont}
\usepackage{multirow}

\usepackage{array}
\usepackage{makecell}

\usepackage{float}

\begin{document}

%\maketitle

\thispagestyle{empty}
\vspace{5mm}
%\vspace{5cm}

\begin{center}

\def\thefootnote{\fnsymbol{footnote}}
\vspace*{2cm}
{\Large \bf On the 
Vacuum Structure of the $\mathcal{N}=4$ Conformal Supergravity\\
\vspace{0.25cm}	
}
\vspace{1.5cm}
Ioannis Dalianis{\large  $^{a,b}$, Alex Kehagias$^{a}$, Ioannis Taskas$^{a}$ and  George Tringas$^{a}$}
\\[0.5cm]

\vspace{.3cm}
{\normalsize $^{a}${\it   Physics Division, National Technical University of Athens, \\15780 Zografou Campus, Athens, Greece}}\\

\vspace{15pt}

{\normalsize $^{b}${\it Department of Physics, University of Athens, University Campus, \\15784 Zografou, Greece}}

%\vspace{.3cm}
%{\normalsize { \it $^{b}$ Department of Theoretical Physics and Center for Astroparticle Physics (CAP)\\ 24 quai E. Ansermet, CH-1211 Geneva 4, Switzerland}}\\

\vspace{.3cm}
%{\normalsize { E-mail: dalianis@mail.ntua.gr,kehagias@central.ntua.gr, ...@mail.ntua.gr and georgiostringas@mail.ntua.gr }}

%\vspace{.2cm}

\end{center}

\vspace{3cm}

\hrule \vspace{0.3cm}
{\small  \noindent \centerline{\textbf{Abstract}}} \\[0.2cm]
\noindent 
We consider ${\cal N}=4$ conformal supergravity with an arbitrary holomorphic function of the complex scalar $S$ which parametrizes the $SU(1,1)/U(1)$ coset. Assuming
 non-vanishings vevs for $S$ and the scalars in a symmetric matrix  $E_{ij}$ of the $\overline{\bf 10}$ of $SU(4)$ R-symmetry group, we  determine the vacuum structure of the theory.  
 We find that the possible vacua  are classified by the  number of zero eigenvalues of the scalar matrix and the spacetime is either Minkowski, de Sitter or  anti-de Sitter. We determine the spectrum of the scalar fluctuations and we find that it contains tachyonic states which however can be removed by appropriate choice of the unspecified at the supergravity level holomorphic function.  Finally,  we also  establish that $S$-supersymmetry is always broken whereas $Q$-supersymmetry exists only on flat Minkowski spacetime.

%We consider ${\cal N}=4$ conformal supergravity with non-vanishing scalars and an arbitrary holomorphic function of the complex scalar which parametrize the $SU(1,1)/U(1)$ coset. In particular, we consider vacua of the theory where the scalars, which  form a matrix in the {\bf 10} of $SU(4)$ and the complex scalar
%have  non-zero vev. We find that the possible vacua of the theory are Minkowski, de Sitter and anti-de Sitter backgrounds which are classified according to the number of zero eigenvalues of the scalar matrix.    We also find that $S$-supersymmetry is always broken whereas $Q$-supersymmetry exists only on flat Minkowski spacetime. 

\vspace{0.5cm} 
 \hrule
\vskip 1cm

\def\thefootnote{\arabic{footnote}}
\setcounter{footnote}{0}
%\maketitle

%\date{\today}

\baselineskip= 12pt

\newpage 
\tableofcontents
\baselineskip=15pt

\section{Introduction}

Conformal supergravity is the supersymmetric completion of conformal or Weyl gravity, described by the Weyl square term. It is invariant under the full superconformal group, which is the supergroup $SU(2,2|\mathcal{N})$, the real form of $SL(4|\mathcal{N})$, where $\mathcal{N}$ counts the number of supersymmetries. This number  cannot be larger than four ($\mathcal{N}\leq 4$) since otherwise, among others, the theory will contain higher spin fields \cite{dF}.  The bosonic part of the above supergroup is $SO(2,4)\times U(\mathcal{N})$ if $\mathcal{N}\neq 4$ or $SO(2,4)\times SU(4)$ if $\mathcal{N}=4$, whereas the fermionic generators are in the $(4,\mathcal{N})+(\bar 4,\bar{\mathcal N})$ for any $\mathcal{N}$. In other words, the superconfromal group contains the standard generators of the conformal group (rotations $M$, translations $P$, conformal boosts $K$ and dilations $D$) as well as the usual $Q$ and the special  $S$ supersymmetry. 

Conformal supergravity employs the Weyl multiplet which is a unique off-shell multiplet and has fewer fields that the corresponding Poincar\'e gravity multiplet. The reason is the high degree of symmetry since the 
local Weyl symmetry implies that certain modes should be absent \cite{FT,FvP}.  
Four dimensional conformal supergravity actions were known for a long time for $\mathcal{N}<4$ \cite{F1,KvN0,KvN,dWvHvP,Berg1,Buch,Tseyt} and the full off-shell action for the $\mathcal{N}=4$ has been also recently  found \cite{Ciceri,Butter1,Butter2}.
Part of the bosonic sector of the theory has previously been obtained in \cite{Tseyt} by utilizing  the conformal anomaly of $\mathcal{N}=4$ vector multiplets.

Conformal supergravity can also be obtained as the massless limit 
$m \to 0$ of the supersymmetric completion of  $m^2 R + Weyl^2$ gravity \cite{FKL1,FKL2,FKL3} (see also \cite{Anselmi1,Donoghue,Salvio}). In general, although such theories contain ghost propagating states \cite{Stelle,Maldacena,Alv}, they are interesting as they arise in twistor-string theory  via closed strings or gauge singlet open strings \cite{BW}.
It is interesting that in the $m\to 0$ limit, the spectrum is re-organized  so that the symmetry is enhanced from super Poincar\'e to superconformal, whereas, at the same time R-symmetries get promoted to  local gauge symmetries.  Let us notice that there are higher curvature supergravities with physical spectrum \cite{FK2,KLO,Ketov,FKR,FKP,Mishra} and a rich vacuum structure \cite{Dalianis:2014aya}.

A particular feature of conformal supergravity is that, although the $\mathcal{N}<4$ theories are unique, the $\mathcal{N}=4$ one is not. In fact, the $\mathcal{N}=4$ theory contains a dimensionless scalar $\phi^\alpha, ~(\alpha=1,2)$ that parametrize the coset $SU(1,1)/U(1)$.  The $U(1)$ is realized as a local symmetry. The corresponding gauge field is composite  with chiral action on the fermions and therefore, the  R-symmetry group is enhanced to $SU(4)\times U(1)$. 

Then, the Weyl square term  among others  in the supersymmetric action,
 is multiplied by a holomorphic function $\cal{H}(\phi^\alpha)$. The existence of this ambiguity of the $\mathcal{N}=4$ conformal supergravity was known for some time \cite{FT2} and it has been explicitly worked out in \cite{Butter1,Butter2}.
 In this work our  aim  is  to  uncover the  vacuum  structure  of  the ${\cal N}=  4$  conformal  supegravity  where  scalar  fields  are also excited 
  looking for maximally symmetric background solutions.
 
The structure of the paper is as follows: In the next section 2, we describe the spectrum of the $\mathcal{N}=4$ conformal supergravity and the corresponding action. In section 3, we explore the vacuum structure of the $\mathcal{N}=4$ theory and finally, we conclude in section 4. 

\section{Spectrum and Action}

In order to establish notation, let us recall the spectrum of the $\mathcal{N}=4$ confromal supergravity. Greek letters $\mu,\nu,\ldots$ denote  space-time indices, $a,b,\ldots$ denote  tangent space indices and $i,j,\ldots$ are $SU(4)$ indices. The bosonic sector contains the vierbein $e_\mu^a$, the $SU(4)$ gauge field ${V_{\mu}}^i_j$ and the gauge field $b_\mu$ which gauges the dilatations.  There are also  composite gauge fields describing the spin connection ${\omega_\mu}^a_b$,
the gauge field $f_\mu^a$ associated to conformal boosts and the composite $U(1)$ gauge field $A_\mu$. The bosonic sector is completed with a complex 
anti-self-dual tensor field ${T_{ab}}^{ ij}$ which is in the ${\bf 6}$  of $SU(4)$, the  complex scalars $E^{ij}$ in the ${\bf 10}$ and the auxiliary pseudoreal  scalars ${D_{ij}}^{kl}$ in the ${\bf 20'}$ of $SU(4)$. Finally, the bosonic sector is completed  by the scalars $\phi^\alpha ~(\alpha=1,2)$ which parametrize 
the coset $SU(1,1)/U(1)$. They are invariant under dilatations and transform as a doublet under $SU(1,1)$ global transformations. The conditions 
 and the constraints these fields satisfy are\footnote{ We use the notation
\begin{align}
\epsilon_{ab}=
\begin{pmatrix}
0 & -1 \\
+1 & 0 
\end{pmatrix}
\; , \;\;\;
\epsilon^{ab}=
\begin{pmatrix}
0 & +1 \\
-1 & 0
\end{pmatrix}
~~~ \mbox{and} ~~~ 
\eta^{ab}=
\begin{pmatrix}
1 & 0 \\
0 & -1
\end{pmatrix}
\nonumber
\end{align}
}
\begin{eqnarray}
&&{T_{ab}}^{ ij}=-{T_{ba}}^{ ij}=-{T_{ab}}^{ ji}, ~~~{T_{ab}}^{ ij}=-\frac{1}{2}
{\epsilon_{ab}}^{cd}{T_{cd}}^{ ij}, \nonumber \\
&& {D^{ij}}_{kl}=\frac{1}{4} \epsilon^{ij mn}\epsilon_{klpq}
{D^{pq}}_{mn}, ~~~~E^{ij}=E^{ji},~~~~E_{ij}=\big(E^{ij}\big)^*,\ \nonumber \\
&&\phi^\alpha\phi_\alpha=1, ~~~\phi_1=(\phi^{1})^*,~~~\phi_2=-(\phi^{2})^*.
\end{eqnarray}
%We also use the notation 
%\begin{eqnarray}
%{T_{ab ij}}=\big({T_{ab}}^{ ij}\big)^*, ~~~~~E_{ij}=\big(E^{ij}\big)^*. 
%\end{eqnarray}

The fields of the $\mathcal{N}=4$ conformal supergravity are completed by the positive chirality fermions which are the gravitini $\psi_\mu^i$, the S-supersymmetry composite $\phi_\mu^i$  
 and the two spinor
fields, 
$\Lambda_i$ in the ${\bf 4}$ and ${\chi^{ij}}_k$ in 
the ${\bf 20}$ of $SU(4)$. The supersymmetry transformations of the $\mathcal{N}=4$ fields can be found in \cite{Berg1}. 

The full off-shell action of the $\mathcal{N}=4$ conformal supergravity has been constructed in \cite{Butter2}. The pure gravitational part contains the Weyl contribution
\begin{eqnarray}
e^{-1}{\cal L}
%\supset
= \frac{1}{2}\big({\cal H}+\overline {\cal H}\big)W^{\mu\nu\kappa\lambda}W_{\mu\nu\kappa\lambda}+\cdots, \label{we}
\end{eqnarray}
where ${\cal H}$ is a holomorphic function of $\phi^\alpha$. The equations of motion for vanishing fields ${T_{ab}}^{ij}, ~E^{ij},~{D^{ik}}_{kl}$ and constant $\phi^\alpha$ are 
\begin{eqnarray}
B_{\mu\nu}=0, ~~~~\big(\partial_\alpha {\cal H}+\partial_{\bar \alpha}\overline {\cal H}\big)W^{\mu\nu\kappa\lambda}W_{\mu\nu\kappa\lambda}=0,
\label{eom}
\end{eqnarray}
where $\partial_\alpha=\partial/\partial \phi^\alpha$, $\partial_{\bar \alpha}=\partial/\partial \phi^{\bar \alpha}$ and 
\begin{eqnarray}
B_{\mu\nu}=\nabla^\rho\nabla_\sigma {W^\sigma}_{\mu\rho\nu}+\frac{1}{2}
R^{\rho\sigma}W_{\rho\mu\sigma\nu}  \label{bach}
\end{eqnarray}
is the Bach tensor. Clearly, conformally flat backgrounds are solutions of the equations of motion Eq.(\ref{eom}). In particular, Minkowski, de Sitter and Anti-de Sitter spacetimes are maximally symmetric vacuum solutions. These solutions are however, trivial in the sense that they do not involve any field other than the vierbein and are indistinguishable in the Weyl theory (they are all maximally symmetric and have vanishing Weyl tensor). Our aim here is to uncover (part of)   the vacuum structure of the $\mathcal{N}=4$ conformal supegravity where scalar fields are also excited. Since we are looking for maximally symmetric backgrounds we will only assume non-vanishing scalars $E^{ij}$ and $\phi^\alpha$, since a non-vanishing tensor field ${T_{ab}}^{ij}$ will in general reduce the background symmetry. In this case, the relevant bosonic part of the action, in the $b_\mu=0$ gauge, is 
\begin{align}
 \mathcal{L} =& +\mathcal{H} \Big[ \frac{1}{2}W^{\mu\nu\rho\sigma}W_{\mu\nu\rho\sigma} +\frac{1}{4}E_{ij}D_\mu D^\mu E^{ij}
 +\frac{1}{8}D^{ij}_{\;\;\;kl}D^{kl}_{\;\;\;ij} -\frac{1}{16}E_{ij}E^{jk}E_{kl}E^{li} + \frac{1}{48}(E_{ij}E^{ij})^2 \Big] \nonumber \\
& +\mathcal{DH}\Big[ +\frac{1}{16}D^{ij}_{\;\;\;kl}(E_{im}E_{jn}\epsilon^{klmn}) \Big] 
%\\&
+\frac{\mathcal{D}^2\mathcal{H}}{384}E_{ij}E_{kl}E_{mn}E_{pq}\epsilon^{ikmp}\epsilon^{jlnq} + h.c.\, , \label{ac1}
\end{align}
where $D_\mu$ is the (super)conformal covariant derivative. 
The fields $D^{ij}_{\;\;kl}$ are auxiliaries as they appear only algebraically  in Eq.(\ref{ac1}). Integrating them out by using their equation of motion 
%They turn out to be 
\begin{align}
    D^{ij}_{\;\;\;kl} = -\frac{1}{4}\frac{\mathcal{DH}}{\mathcal{H}}E_{km}E_{ln}\epsilon^{ijmn}, \label{aux}
\end{align}
%and integrating them out,
we find that the 
lagrangian in Eq.(\ref{ac1}) is written as 
\begin{align}
 \mathcal{L} =& \mathcal{H} \Big[\frac{1}{2}W^{\mu\nu\rho\sigma}W_{\mu\nu\rho\sigma}-\frac{1}{4}\nabla_\mu E_{ij}\nabla^\mu E^{ij}-\frac{1}{24}RE_{ij}E^{ij}- \frac{1}{16}E_{ij}E^{jk}E_{kl}E^{li} + \frac{1}{48}(E_{ij}E^{ij})^2 \Big]  \nonumber\\
& -\frac{1}{64}\frac{(\mathcal{DH})^2}{\mathcal{H}}E_{ij}E_{kl}E_{mn}E_{pq}\epsilon^{ikmp}\epsilon^{jlnq} + \frac{\mathcal{D}^2\mathcal{H}}{384}E_{ij}E_{kl}E_{mn}E_{pq}\epsilon^{ikmp}\epsilon^{jlnq} + h.c. \label{final}
\end{align}
Note that apart from the usual Weyl square term, the scalars $E_{ij}$ are conformally coupled to the curvature in the standard way. However, in order gravity to be attractive in the infrared,  vector multiplets coupled to the supergravity multiplet are needed \cite{deroo1,deroo2,FKL3}.

We are interested in maximally symmetric vacuum solutions (Minkowski, de Sitter or anti-de Sitter) so that 
\begin{eqnarray}
R_{\mu\nu\rho\sigma}=\Lambda \big(g_{\mu\rho}g_{\nu\sigma}-g_{\mu\sigma}g_{\nu\rho}\big)~, ~~~~E_{ij}=\mbox{const.}~, ~~~\phi^\alpha=\mbox{const.}~,
\end{eqnarray}
where $\Lambda$ is the cosmological constant which, in our conventions, it is positive, negative or zero for de Sitter, anti-de Sitter or Minkowski spacetime, respectively.
In this case, the equations of motion which follow from the Lagrangian of Eq.(\ref{final}) are 
\begin{eqnarray}
&&A B_{\mu\nu} +C\left(R_{\mu\nu}-\frac{1}{2}g_{\mu\nu}R\right)+\frac{1}{2}Vg_{\mu\nu}=0~, \label{eom1}\\
&&R\partial_I C-\partial_I V=0~, \label{eom2}
\end{eqnarray}
%where $B_{\mu\nu}$ is the Bach tensor and in Eq.(\ref{eom2}), the index $I$ enumerates collectively  all scalar fields. We have defined the following functions
where the index $I$ enumerates collectively  all scalar fields and $B_{\mu\nu}$ is the Bach tensor defined in Eq.(\ref{bach}). 
We have also introduced the functions
\begin{eqnarray}
&&A(\phi^{\alpha},\phi_{\alpha})= \mathcal{H} + \mathcal{\overline{H}}~, ~~~~~ C(\phi^{\alpha},\phi_{\alpha},E_I)=-\frac{1}{24}A \, E_{ij}E^{ij}~, \label{AB} \nonumber \\
&&G(\phi^{\alpha},\phi_{\alpha})= \Big(\frac{(\mathcal{DH})^2}{\mathcal{H}}-\frac{\mathcal{D}^2\mathcal{H}}{6}\Big) \label{G}~,
\end{eqnarray}
which, for shorthand, we will refer to them simply as  $A$, $C$ and $G$. The scalar potential $V$ turns out to be
\begin{align}
V=& \mathcal{H} \Big( \frac{1}{16}E_{ij}E^{jk}E_{kl}E^{li} - \frac{1}{48}(E_{ij}E^{ij})^2 \Big) \nonumber \\& +\frac{1}{64}\Big(\frac{(\mathcal{DH})^2}{\mathcal{H}}-\frac{\mathcal{D}^2\mathcal{H}}{6}\Big)E_{ij}E_{kl}E_{mn}E_{pq}\epsilon^{ikmp}\epsilon^{jlnq} + h.c. \label{potential}
\end{align}
Since the Bach tensor vanishes for conformally flat geometries as the ones we are after (Minkowski, de Sitter or anti-de Sitter), 
Eqs. (\ref{eom1}) and (\ref{eom2}) are simply 
\begin{eqnarray}
\Lambda=\frac{V}{2C}~, ~~~~~~2V \partial_I C-C \partial_I V=0~. \label{ee1}
\end{eqnarray}
The first equation of Eq.(\ref{ee1}) specifies the cosmological constant $\Lambda$ and the second one can be written as
\begin{eqnarray}
\frac{C^2}{V}\partial_I\frac{V}{C^2}=\frac{1}{V_{eff}}\partial_I V_{eff}=0~.  \label{veff}
\end{eqnarray}
 Therefore, the vacua of the theory are  specified by  the extrema of the ``effective" potential 
\begin{align}
V_{eff}=\frac{V}{C^2}~. \label{effective}
\end{align}
Our aim in the following is to determine these extrema.

\section{Vacuum solutions}
In order to find the vacuum structure we should minimize  the effective scalar potential in Eq. (\ref{effective}), which is explicitly written as  
\begin{align}
V_{eff}=\bigg(\frac{24}{\big({\cal H}+\bar {\cal H}\big)E_{ij}E^{ij}}\bigg)^2
&\bigg\{\mathcal{H} \Big( \frac{1}{16}E_{ij}E^{jk}E_{kl}E^{li} - \frac{1}{48}(E_{ij}E^{ij})^2 \Big) \nonumber \\& +\frac{1}{64}\Big(\frac{(\mathcal{DH})^2}{\mathcal{H}}-\frac{\mathcal{D}^2\mathcal{H}}{6}\Big)E_{ij}E_{kl}E_{mn}E_{pq}\epsilon^{ikmp}\epsilon^{jlnq} 
\bigg\}+ h.c. \label{potential}
\end{align}
Clearly, we cannot proceed without specifying the exact form of the holomorphic function ${\cal H}(\phi^\alpha)$. Therefore, we need to choose the explicit form of ${\cal H}(\phi^\alpha)$ as, at this point, it is totally arbitrary, although it is expected to be specified in a more fundamental theory. But before, we need to elaborate on its properties and structure
%we are forced to do the following ansatzes 
which will be discussed in the next sections.
%both for $E_{ij}$ and for the choice of the function $\mathcal{H}(\phi^\alpha)$.

\subsection{Structure of  $\mathcal{H}$}
%As discussed previously, 
%The scalar fields $\phi^{\alpha}$ of  the $\mathcal{N}=4$ Weyl multiplet  parametrize the coset $SU(1,1)/U(1)$.  
The derivatives ${\mathcal{D}}$ on the scalar manifold $SU(1,1)/U(1)$ which appear in the lagrangian Eq. (\ref{ac1}) are defined as \cite{Butter1}
\begin{align}
    & \mathcal{D}=-\phi^{a}\epsilon_{ab}\frac{\partial}{\partial\phi_b}~, ~~~~~~ \overline{\mathcal{D}}=+\phi_a\epsilon^{ab}\frac{\partial}{\partial\phi^b}~, ~~~~~~
    \mathcal{D}^0=\phi^{\alpha}\frac{\partial}{\partial\phi_a}-\phi_{\alpha}\frac{\partial}{\partial\phi_{\alpha}}~. \label{derivatives}
\end{align}
Clearly with the definitions above,  the derivative ${\cal DH} $ of the holomorphic function ${\cal H}(\phi^\alpha)$ is also  holomorphic ($\bar{\cal D}{\cal DH} =0$). 
Now, due to the constraint $\phi^{\alpha}\phi_{\alpha}=1$,  functions of the form $\mathcal{H}(\phi_2/\phi_1)$ are holomorphic. Therefore
it is convenient to  define the fields $S$ and $\psi$ as 
\begin{align}
S = \frac{\phi_2}{\phi_1}~,~~~~ \overline{S} = -\frac{\phi^2}{\phi^1}~, ~~~~ e^{2i\psi}=\frac{\phi^1}{\phi_1}~, \label{sspsi}
\end{align}
In this parametrization, the field  $S$ parametrize the Poincare disk $0\leq\vert S\vert<1$ whereas the phase $\psi$ describes the  $U(1)$. 
 The derivatives Eq.(\ref{derivatives}) are expressed now as 
\begin{align}
   & \mathcal{D}= -e^{+2i\psi} \Big( (1-S\overline{S})\partial_s + \frac{i}{2}\overline{S}\partial_{\psi} \Big)~, \nonumber \\
   &\overline{\mathcal{D}}=(\mathcal{D})^*~,~~~~~ \mathcal{D}^0=-i\partial_{\psi}~.\label{D1}
\end{align}
Then, the operator  $\mathcal{D}^{2}$ turns out to be 
\begin{align}
    \mathcal{D}^{2}= \overline{S}e^{2i\psi}\mathcal{D} + e^{4i\psi}\Big(&-\overline{S}(1-S\overline{S})\partial_S +\frac{i}{2}\overline{S}(1-S\overline{S})\{\partial_S,\partial_{\psi}\} \nonumber \\
    &- \frac{1}{4}\overline{S}^2\partial^2_{\psi} + (1-S\overline{S})^2\partial^2_S \Big) \;.  \label{D2}
\end{align}
One can see that for any choice of $\mathcal{H}(S)$ the first order derivative is also holomorphic $\overline{\mathcal{D}}\mathcal{D}\mathcal{H}=0$.
With the expressions in Eqs. (\ref{D1}) and (\ref{D2}),  we have for the quantities $\mathcal{DH}$ and $\mathcal{D}^{2}\mathcal{H}$  that enter in the lagrangian
\begin{align}\label{example}
    &\mathcal{DH}= -e^{+2i\psi}(1-S\overline{S})\partial_S\mathcal{H}\; , \\
    &\mathcal{D}^{2}\mathcal{H}= -e^{+4i\psi}(1-S\overline{S})\Big(2\overline{S}\partial_S\mathcal{H}-(1-S\overline{S})\partial_S^2\mathcal{H}\Big)\; ,
\end{align}

\subsection{Vacua of the $\mathcal{N}=4$ conformal supergravity}
The scalars $E_{ij}$ are in the $\overline{\mathbf{10}}$ of $SU(4)$ and therefore can be represented as a complex symmetric $4\times 4$ matrix. One of the simplest configuration is the one where $E_{ij}$ is diagonal and takes therefore the form
%In this subsection we start 
%by considering a specific structure for the scalars $E_{ij}$. Namely, we will assume that the 
%the following two general ansatzes trying to classify the effective vacuum solutions, 
%the cosmological constant and then the symmetry breaking pattern of $SU(4)$. The general ansatz for scalar matrix 
%scalar matrix $E_{ij}$ is diagonal, and in particular of the form
\begin{align}
E_{ij} =
%\; : \;\;
\begin{pmatrix}
E_1 & 0 & 0 & 0 \\
0 & E_2 & 0 & 0 \\
0 & 0 & E_3 & 0 \\
0 & 0 & 0 & E_4
\end{pmatrix} =E_i \delta_{ij}~~~~(\mbox{no summation})\; . \label{eij}
\end{align}
In order to  proceed now, we will distinguish two different cases for the holomorphic function ${\cal H}$:
%
%study the properties of the vacuum when the holomorphic function ${\cal H}$ according to the two cases: 
1) ${\cal DH}={\cal D{H}}_0=0$, and 2) ${\cal DH}\neq const$.
We will examine these cases separately below.
\subsubsection{Constant holomorphic function}\label{subsectionforHconst}
The first case corresponds to a constant homolophic function ${\cal H}$. We can take here  $\mathcal{H}(S)\equiv \mathcal{H}_0=\mbox{const.}$, so that 
\begin{eqnarray}
A={\cal H}_0+\overline{\cal H}_0=\frac{1}{\alpha^2}~, ~~~~~~ C=-\frac{1}{24\alpha^2} E_{ij}E^{ij}~, ~~~~~~~G=0~.
\end{eqnarray}
Then,  the effective potential in  Eq.(\ref{potential}) and 
%from Eqs.(\ref{ee1}) and (\ref{effective}) one can see that 
the cosmological constant in Eq.(\ref{ee1}) turn out to be 
\begin{eqnarray}
V_{eff}=-12 \alpha^2\frac{\big(\sum_i\vert E_i\vert^2\big)^2-3\sum_i\vert E_i\vert^4}{\big(\sum_i\vert E_i\vert^2\big)^2}~, 
\end{eqnarray}
and 
\begin{eqnarray}
\Lambda=+\frac{1}{4}\frac{\big(\sum_i\vert E_i\vert^2\big)^2-3\sum_i\vert E_i\vert^4}{\sum_i\vert E_i\vert^2} \label{cosmo1}
\label{v00}
\end{eqnarray}
for $i=1...4$, respectively. Note that, unlike the effective potential, the cosmological constant does not depend on the overall factor $A$ as it should  for a constant holomorphic ${\cal H}$. 
The effective potential, $V_{eff}$ is  minimized at several points with different gauge symmetry breaking patterns and different values of the cosmological constant reported below. 

The extrema of the effective potential are solution of Eq. (\ref{veff}) and they can be classified according to the number of zero eigenvalues of the scalar-field matrix $E_{ij}$. 
%The derivative of the effective potential with respect to the $E_{ij}$ fields when vanishes gives the extrema of the potential. 
These extrema fall in the following classes:
\begin{enumerate}
 \item [$a)$]
The first class consists of configurations of  only one non-zero  eigenvalue of $E_{ij}$ (let say $\vert E_1\vert =\vert E_2\vert =\vert E_3
\vert =0$ and 
$\vert E_4\vert =\vert E\vert\neq 0$). The associated cosmological constant is negative 
\begin{align}
\Lambda=-\frac{\vert E\vert^2}{2} ~,  \label{ads2}
\end{align}
giving rise to an anti-de Sitter background, an unbroken $SU(3)$ and positive effective potential $\langle V_{eff}\rangle=+24\alpha^2.$
   
\item [$b)$]
A second class of  solutions is when two eigenvalues of the matrix are zero (let say $\vert E_1\vert =\vert E_2\vert=0$) while the others have equal modulus ($\vert E_{3}\vert=\vert E_{4}\vert=\vert E\vert$). This breaks $SU(4)\rightarrow SU(2)$  and  the cosmological constant turns out to be again negative
\begin{align}
  \Lambda=-\frac{\vert E\vert^2}{4}~,  \label{ads1}
\end{align}
corresponding to an anti-de Sitter background. The effective potential is positive in this case $\langle V_{eff}\rangle=+6\alpha^2$.

\item [$c)$]
A third class of solutions is obtained when there is a single zero eigenvalue (let say $\vert E_1 \vert=0$) and the  modulus of the other eigenvalues are equal ($\vert E_2\vert=\vert E_3\vert=\vert E_4\vert$). 
%These relations respect cyclic permutation over the indices.
In this case, it can easily be verified that the cosmological constant in Eq.(\ref{v00}) vanish 
\begin{eqnarray}
\Lambda=0~,  \label{mink}
\end{eqnarray}
corresponding to a  Minkowski vacuum and an unbroken $U(1)$ symmetry.

\item [$d)$]
A final class of extrema of the effective potential contains scalars $E_{ij}$ with non-zero eigenvalues but with equal modulus ($|E_1|=|E_2|=|E_3|=|E_4|=|E|)$.
The cosmological constant is positive in this case and turns out to be
\begin{align}
\Lambda=+\frac{\vert E\vert^2}{4}~, \label{ads}
\end{align}
corresponding to a de Sitter background 
whereas the $SU(4)$ symmetry is in generally broken. However,  when the fields $E_i$ ($i=1,2,3,4$) are equal and not only their modulus, the $SU(4)$ is broken down to 
$O(4)$.  The effective energy turns out to be  $\langle V_{eff}\rangle=-3\alpha^2$.

\end{enumerate}

The number of zero eigenvalues, the vacuum energy and the symmetry breaking for each case are collected in the following table:
\begin{center}\label{table1}
\begin {tabular}{|c|c|c|c|c|}
     \hline 
     \thead{Vacuum \\ case } & \thead{Zero \\ eigenvalues } & \thead{VEV of effective\\ potential} & \thead{Background\\ \phantom{x}} & \thead{Unbroken subgroup \\ of SU(4)}   \\ 
     \hline \hline
     a & 3 &  \makecell{$\langle V_{eff}\rangle$ $>$ 0}  & AdS & SU(3)  \\
     \hline
    b & 2 &  \makecell{$\langle V_{eff}\rangle$ $>$ 0}  & AdS & SU(2)   \\
     \hline
    c & 1 &  \makecell{$\langle V_{eff}\rangle$ $=$ 0}  & Minkowski & U(1)   \\
     \hline
    d & 0 &  \makecell{$\langle V_{eff}\rangle$ $<$ 0}  & dS & completely broken   \\
     \hline
\end{tabular}
\captionof{table}{\label{CCtable}The vacua of the theory in the case of a diagonal scalar matrix  $E_{ij}$ are collectively presented in this table.  As we will see later, most of the vacua are non-supersymmetric.}
\end{center}

\subsubsection{Non-constant holomorphic function}
Let us now examine the effective potential in Eq.(\ref{potential}) when the holomorphic function ${\cal H}$ is non-constant. 
In this case, the effective potential for a diagonal scalar-field matrix $E_{ij}$ of the form (\ref{eij}) turns out to be 
\begin{align}
    V_{eff}=-\frac{12}{A(S,\overline{S})}\frac{\big(\sum_i\vert E_i\vert^2\big)^2-3\sum_i\vert E_i\vert^4}{\big(\sum_i\vert E_i\vert^2\big)^2}+ \frac{216}{A(S,\overline{S})^2}\Bigg\{\frac{\prod_{i}E_i\times G}{\big(\sum_i\vert E_i\vert^2\big)^2} + h.c. \Bigg\} \label{potentialHD} ~~.
\end{align}
Comparing the above potential to the one of the previous section (constant ${\cal H}$), we see that the first term is the same but the function $A\equiv A(S,\overline{S})$ is not constant anymore. The extra contribution in the brackets appears due to the presence of the derivatives of the holomorphic function $\mathcal{H}$. It is important to notice that since the second term in the effective potential in Eq.(\ref{potentialHD})  is a product of the eigenvalues of the scalar matrix $E_{ij}$, it vanishes when the determinant of $E_{ij}$ is zero, i.e., when at least one of its eigenvalues vanishes.

We will determine now the cosmological constant and compare it to Eq.(\ref{cosmo1}) of the previous section. The effective potential in Eq.(\ref{potentialHD}) has schematically the form
\begin{align}
    V_{eff}=\frac{1}{A(S,\overline{S})}V^{H}+\frac{1}{A(S,\overline{S})^2}V^{DH}~~,
\end{align}
where $V^{H}$ and $V^{DH}$ can be read off from Eq.(\ref{potentialHD}) to be 
\begin{align}
   & V^{H}=-12\frac{\big(\sum_i\vert E_i\vert^2\big)^2-3\sum_i\vert E_i\vert^4}{\big(\sum_i\vert E_i\vert^2\big)^2}~~, \\
   &V^{DH}=216\Bigg\{\frac{\prod_{i}E_i\times G}{\big(\sum_i\vert E_i\vert^2\big)^2} + h.c. \Bigg\} \label{CDH} ~~.
\end{align}
Then, from Eqs.(\ref{AB}) and (\ref{ee1}) we see that the cosmological constant can always be written as
\begin{align}
    \Lambda&=-\frac{V_{eff}}{48}A(S,\overline{S})\sum_{i}\vert E_{i}\vert^2\nonumber \\
    &=
    -\frac{1}{48}\Big(V^{H}+\frac{1}{A(S,\overline{S})}V^{DH} \Big)\sum_{i}\vert E_{i}\vert^2 \label{cosmoHD} ~.
\end{align}
When the  derivatives of ${\cal H}$ vanish as what the case in the previous section, the cosmological constant depends only on the vevs of the $E_{ij}$ fields as in the case with constant holomorphic function ${\cal H}$. The
derivative terms just add an extra contribution which depends on the vev of the $S$ field which parametrize the manifold $SU(1,1)/U(1)$.

\subsubsection{Vacua for non-constant (general) holomorphic function}\label{subsectionforHnonconst}
 
For non-constant holomorphic function and proceeding as before, the non-trivial critical points of  the effective potential in Eq.(\ref{potentialHD}) turns out to be 
(with $i,j,k,l=1,2,3,4$
and $i\neq j\neq k\neq l$)

\begin{align}
%\label{eigen0}
& a)~~~\vert E_i\vert=\vert E_j\vert=\vert E_j\vert=0~, ~~~~\vert E_k\vert=E\neq 0\label{s1}\\
 &b)~~~\vert E_i\vert =\vert E_j\vert\neq 0~, ~~~~~\vert E_k\vert =\vert E_l\vert=0~ .\label{s2}\\
   & c)~~~E_i \vert E_i\vert^2=\frac{A}{3G}E_j^*E_k^*E_l^* ~,~~~~\mbox{or} \nonumber   \\
   & \phantom{5)}~~~E_i \vert E_i\vert^2=\frac{27 G^*\vert G\vert^2}{A^3}E_j^*E_k^*E_l^* ~~~~\mbox{and} ~~~E_j \vert E_j\vert^2=\frac{A}{3G}E_i^*E_k^*E_l^*~, \label{s3} \\
  % &\text{or} \nonumber \\
   & d)~~~E_i \vert E_i\vert^2=\left(\frac{G^*}{G}\right)^{1/2}E_j^*E_k^*E_l^*, \label{s4}
\end{align}
where, the functions $A,G$ are evaluated at the $S$-critical points. The latter can be determined whenever the explicit form of ${\cal H}(S)$ is known. 
At the above points of Eqs. (\ref{s1})-(\ref{s4}), the vacua can be classified according to the number of zero eigenvalues of $E_{ij}$ as follows:

\begin{enumerate}
    \item [$a)$] With three  zero and one non-zero eigenvalue $E$, corresponding to Eq. (\ref{s1}),  the cosmological constant turns out to be 
\begin{align}\label{vac2}
     \Lambda=-\frac{1}{2}\vert E\vert^2~,
\end{align}
and the effective potential is 
%$   V_{eff} =\frac{24}{A(S,\overline{S})}$.
\begin{align}
  V_{eff}=\frac{24}{A(S,\overline{S})}~. \label{eff11}  
\end{align}

\item [$b)$] For two zero   (let say $E_1=E_2=0$) and two non-zero eigenvalues of equal modulus  ($\vert E_3\vert=\vert E_4\vert=\vert E\vert$), corresponding to Eq. (\ref{s2}),
the cosmological constant is 
\begin{align}\label{vac8}
     \Lambda=-\frac{1}{4}\vert E\vert^2~,
\end{align}
with  
\begin{align} 
   V_{eff}=\frac{6}{A(S,\overline{S})}~. \label{eff22} 
\end{align}
As dictated by Eq. (\ref{cosmoHD}) the cosmological constants found in these last two classes of solutions are  equal to those of Eq. (\ref{ads2}) and (\ref{ads1}) respectively.

\item [$c)$]
For non-zero eigenvalues which satisfy the relations in Eq.(\ref{s3}), the effective potential and the cosmological constant have the form 
\begin{align}
    V_{eff}=\frac{72}{A(S,\overline{S})}\frac{\vert G\vert^2}{A(S,\overline{S})^2+3\vert G\vert^2}~, ~~~~~~ \Lambda=-\frac{3}{2}\frac{\vert G\vert^2}{A(S,\overline{S})+3\vert G\vert^2}\sum_i\vert E_i\vert^2~.
\end{align}
In this case the effective potential is always positive and the cosmological constant is always negative. Since the derivatives of ${\cal H}$ also shape the vacuum structure, it is interesting to study them explicitly using the expression in Eq.(\ref{G}) and the derivative operator in Eq.(\ref{example})
\begin{align}
   G=e^{4i\psi}(1-\vert S\vert^2)\Big(\frac{(1-\vert S\vert^2)\mathcal{H}'^2}{\mathcal{H}}+\frac{2\overline{S}\mathcal{H}'-(1-\vert S\vert^2)\mathcal{H}''}{6}\Big)~, \label{GG}
\end{align}
where the prime denotes partial derivative with respect to the $S$ field. The first part of this function is always positive definite since the 
target space of $S$ is the Poincare disk, while the sign of the second part depends on the choice of the holomorphic function. One can require the function $G$ to vanish or only the second part to vanish which leads to a second order differential equation for $\mathcal{H}(S)$. Solving both these cases the solution is a non-holomorphic function thus neither the second part nor the whole $G$ can vanish with a proper selection of $\mathcal{H}(S)$. 
\item  [$d)$]
Finally, if eigenvalues that satisfy Eq.(\ref{s4}), the effective potential and cosmological constant turn out to be
\begin{align}
    ~~~~~~~~~~~V_{eff}=-\frac{3}{A(S,\overline{S})}\mp\frac{27\vert G\vert}{A(S,\overline{S})^2}~, ~~~~~~ \Lambda=\frac{1}{16}\Big( 1 \pm \frac{9\vert G\vert}{A(S,\overline{S})}\Big)\sum_i\vert E_i\vert^2~. \label{SOLS4}
\end{align}
Whether the cosmological constant is zero, positive or negative depends on the values of $G$ and $A$. In the first case, corresponding to the plus sign in Eq.(\ref{SOLS4}),  the cosmological constant is always positive. In the second case, corresponding to the minus sign in Eq.(\ref{SOLS4}), there are three possibilities according to the value of $\lambda\!=\!A-9\vert G\vert$~:   anti-de Sitter for $\lambda<1$, de Sitter for $\lambda>1$ and Minkowski for $\lambda=1$. 
\end{enumerate}

We sum up our results in the following table:

\begin{center}
\begin{tabular}{|c|c|c|c|c|c|}
     \hline 
\thead{Vacuum \\ case } &     \thead{Number of zero \\ eigenvalues } & \thead{VEVs of \\ $E_{ij}$} & \thead{Cosmological \\Constant $\Lambda$} & \thead{Symmetry Breaking \\ of SU(4)}  \\
     \hline\hline
     a &  3 &  \makecell{$\langle V_{eff}\rangle$ $>$ 0}  & AdS & SU(3)  \\
     \hline
     b & 2 &  \makecell{$\langle V_{eff}\rangle$ $>$ 0}  & AdS & SU(2)  \\
     \hline
    c & 0 &  \makecell{$\langle V_{eff}\rangle$ $>$ 0}  & AdS & completely broken  \\
     \hline
    d & 0 &  \makecell{ $\langle V_{eff}\rangle$ $<$ 0 \\ $\langle V_{eff}\rangle$ $=$ 0 \\ $\langle V_{eff}\rangle$ $>$ 0}  & \makecell{ dS \\ Minkowski \\ AdS } & completely broken  \\
     \hline
\end{tabular}
\captionof{table}{\label{CC2table}In this table, we collectively present the  vacua of the theory, indicating  the number of eigenvalues and their relation to the cosmological constant and the symmetry breaking pattern. The Minkowski vacuum exists only if we fine tune  $\lambda=1$. }
\end{center}

\subsubsection{Explicit examples for non-constant holomorphic function}
So far we have kept the discussion general and have classified the possible vacua according to the number of  eigenvalues that vanish in the vacuum.
In the case where the holomorphic function is constant the effective potential and the cosmological constant are independent of $\mathcal{H}(S)$ (since in this case ${\cal H}$ is an overall coupling constant) and the vacua are defined in Sec.(\ref{subsectionforHconst}). On the other hand, when the holomorphic function is non-constant, the value of the cosmological constant depends on the choice of $\mathcal{H}(S)$ which shapes the vacuum structure in a different way.

As we have discussed above, the function $\mathcal{H}(S)$ is arbitrary and is expected to be specified in a more fundamental theory.
However, in order to be more explicit and for illustrative purposes, we will explore here  some explicit examples with different forms of the function $\mathcal{H}(S)$. For this, we have to distinguish the possible vacua into two groups, the group I which contains the cases $a),$ and $b)$  
and the group II which contains the cases $c),$ and $d)$. The reason is that the effective potential in group I is determined entirely in terms of the function $A$, whereas the effective potential for group II is determined from both functions $A$ and $G$.  
We start from the vacua I in Eqs.(\ref{vac2}) and (\ref{vac8}) where the cosmological constant does not depending on the $S$ field. If we choose the holomorphic function to be linear
\begin{equation}
    \mathcal{H}(S)=S~,
\end{equation}
it is obvious that the effective potential 
\begin{align}
    V_{eff}(S,\overline{S})=\frac{3}{Re\, S}~,
\end{align}
has a runaway behaviour and no critical points. 
Critical point of the potential exist only when the function $\mathcal{H}(S)$ has critical points itself.
As a particular example for 
\begin{align}
\mathcal{H}(S)=\pm S^2+S~,    
\end{align}
the effective potential has its extrema at $S_0=\mp\frac{1}{2}$ for  the solution in Eq.(\ref{vac8})
\begin{align}
    \langle V_{eff}\rangle=\frac{6}{\mp S^2+S+h.c.}\Big\vert_{S_0=\overline{S}_0}=\mp12,~~~~~~ \Lambda=-\frac{\vert E_3\vert^2}{2}~.
\end{align}
Note that   the critical points of the effective potential should lie inside the Poincare disk $0\leq \vert S\vert <1$.

Next, we   examine the vacua II where all the eigenvalues of $E_{ij}$  are non-zero. 
For convenience we assume that $E_{ij}=E\delta_{ij}$ and  
%belongs to the class where all the eigenvalues are non-zero and equal and will help us to investigate the vacuum with the presence of non-constant holomorphic ${\cal H}$.
%where the SU(4) is unbroken 
%and examine the case where $\mathcal{H}\neq const$ since the vacuum and the cosmological constant for constant holomorphic function have been discussed above.  
by minimizing the effective potential in Eq.(\ref{potentialHD}) we find 
\begin{align}
& E^*=\pm E\Big(\frac{G}{G^*}\Big)^{1/4} \phantom{i} ~~~~\mbox{with} ~~~~~~~ \Lambda=\frac{1}{4}\Big( 1 + \frac{9\vert G\vert}{A(S,\overline{S})}\Big)\vert E\vert^2~, \label{sol11}\\
& E^*=\pm iE\Big(\frac{G}{G^*}\Big)^{1/4} ~~~~ \mbox{with} ~~~~~~~ \Lambda=\frac{1}{4}\Big( 1 - \frac{9\vert G\vert}{A(S,\overline{S})}\Big)\vert E \vert^2~. \label{sol22}
\end{align}
In the previous subsection we saw that the derivatives of ${\cal H}$, which are contained in the functions $G$ and $G^*$ in Eq.(\ref{G}), give an extra contribution to the effective potential and the cosmological constant. By setting them to zero we arrive to the maximally symmetric solutions independent of the $S$ fields. The critical points agree with the general solutions in Eq.(\ref{s4}).
The solutions in Eqs. (\ref{sol11}) and (\ref{sol22}) belong to the class $d)$   (i.e. Eq. (\ref{SOLS4})) where all the maximally symmetric cases are possible and thus the holomorphic function $\mathcal{H}(S)$ has to be specified in order to find the vacuum. To see how this works and  in order to proceed further, we choose a power-law form for  the holomorphic function $\mathcal{H}(S)=S^n$ as an example. Note that in the case we are discussing, since both functions $A$ and $G$ appear in the effective potential, it is not necessary ${\cal H}$ to have a critical point.  Then, the cosmological constant in Eq.(\ref{sol11}) turns out to be 
\begin{align}
    \Lambda=\frac{1}{4}\Bigg(1+\frac{3}{2}\frac{\sqrt{n^2S^{-2+n}\overline{S}\,^{-2+n}(-1+\vert S\vert^2)^2(-1-5n+(-1+5n)\vert S\vert^2)^2}}{S^n+\overline{S}\,^n} \Bigg)\vert E\vert^2~. \label{cc1}
\end{align}
The value of $S$ is determined by minimizing the effective potential. Indeed, the critical points of the effective potential can be found for integer values of $n$ at $S=S_0=\overline{S}_0$. We find for example that for $n=-1$, there are two critical points $S_0\sim -0.87$ with $\Lambda\sim 0.2 \vert E\vert^2$ and $S_0\sim 0.93$ with $\Lambda\sim 0.28\vert E\vert^2$, for  $n=-2$, 
$\vert S_0\vert \sim 0.97$ with $\Lambda\sim 0.28 \vert E\vert^2$, in general, $S_0$ lies within the Poincar\'e disk for $n\leq 0$.  At these points, the cosmological constant in Eq.(\ref{cc1}) is always positive ($\Lambda>0$) corresponding to a de Sitter background and increases for larger values of $n$. For $n>0$, $S_0$ is outside the Poincar\'e disk and should not be considered. 
Similarly the cosmological constant in Eq.(\ref{sol22}) and for our specific choice of function $\mathcal{H}$ leads to de Sitter backgrounds (for small  negative values of  $n$) and both de Sitter and anti de Sitter solutions (for large negative values of  $n$).

\subsection{Stability}
In order to determine whether the vacua found in the previous sections are stable,   we have to calculate the masses of the fluctuations around these vacua. We consider the simplest case where $\mathcal{H}=const$ and the only scalar fields considered are $E_{ij}$ and their conjugates since this is the case where the masses can be calculated analytically. In the more general case, numerical calculations are necessary. A small perturbation $\delta E_{ij}$ around the vacuum satisfies the equation 
\begin{align}
    \nabla_{\mu}\nabla^{\mu}\delta E_{ij}-\Big(\frac{2}{3}\Lambda+4\frac{\partial^2V}{\partial E^{ij}\partial E_{kl}}\delta E_{kl}+4\frac{\partial^2V}{\partial E^{ij}\partial E^{kl}}\delta E^{kl}\Big) =0~, \label{eq0}
\end{align}
where the second derivative of the potential is calculated on the vacuum and we have used that $R=4\Lambda$. The  20 real degrees of freedom of
$\delta E_{ij}$  corresponding to the ${\bf 10}+\overline{\bf 10}$ fields can be arranged so that Eq. (\ref{eq0}) can be written as 
$\nabla^2\delta E- {\cal M}^2 \delta E=0$. The square of the $20\times 20$  mass matrix  ${\cal M}^2 $ 
%and the second order perturbation gives the mass term $\delta E^{\dagger} \mathcal{M}^2 \delta E$ which is reproduced by the following 
%mass matrix takes the form 
is of the form 
\begin{equation}\label{Matrix}
\mathcal{M}^2=
\begin{pmatrix}
M^2_{ij\overline{kl}} & M^2_{ij kl} \\
 M^2_{\overline{ij}\overline{kl}}&   M^2_{\overline{ij}kl}
\end{pmatrix}~.
%~,
%\end{equation}
%\setcounter{MaxMatrixCols}{20}
%\begin{equation}
%\delta E^{\dagger} =
%\begin{pmatrix}
%  E_{11} &  E_{22} &  E_{33} &  E_{44} &  E_{12} &  E_{13} &  E_{14} &  E_{23} &  E_{24} &  E_{34} & (...)^* 
%\end{pmatrix} ~,
\end{equation}
%where the ellipsis in the parenthesis term stand for the ${\bf 10}$ fields. 
Stability requires the eigenvalues of the matrix ${\cal M}^2$ to be non-zero for Minkowski and de Sitter backgrounds. However, in the case of AdS vacua Eq.(\ref{ads1}) and (\ref{ads2}) the eigenvalues $m^2$ should satisfy  the BF-bound  \cite{BF1,BF2}
\begin{align}\label{BF}
    m^2_{ij}\geq  -\frac{3}{4}\vert \Lambda \vert~. 
\end{align}
%for $m^2_{ij}$ the eigenvalues of the modified mass term, $r^2=3/\Lambda$ the curvature radius and $p=2$ since this relation is derived for %$\text{AdS}_{p+2}$.
%We first present in detail the simplest case of mass calculation which corresponds to the AdS vacuum with one non-zero eigenvalue for $E_{ij}$ in %Eq.(\ref{ads2}).
%The submatrix $M_{ij,\overline{kl}}$ is diagonal
%\begin{align}
 %   M_{ij,\overline{kl}}=\{R,R,R,K,R,R,K,R,K,K\}~,
%\end{align}
%while there exist non-diagonal elements $M_{44,44}=N$ and its conjugate. The entries are given by
%\begin{align}
 %   R&=\frac{1}{6}(\vert E\vert^2+\Lambda)~, \\
  %  K&=\frac{1}{6}(2\vert E\vert^2+\Lambda)~, \\
   % N&=\frac{1}{6} E^2~.
%\end{align}
For the  AdS vacuum obtained from an $E_{ij}$ with one non-zero eigenvalue of Eq.(\ref{ads2}), we find that the following eigenvalues of the mass matrix 
\begin{align}
    &m^2_{11}=\frac{1}{6}(-\vert E\vert^2+4\Lambda)=-\frac{1}{2}\vert E\vert^2~, ~~~~ m=6 ~~~ \label{tachyonAdS1}\\
    &m^2_{22}=\frac{1}{2}m^2_{11}=-\frac{1}{4}\vert E\vert^2 ~, ~~~~~~~~~~~~~~~~ m=6 ~~~ \\
    &m^2_{33}=\frac{1}{6}(2\vert E\vert^2+4\Lambda)=0 ~, ~~~~~~~~~~~~~ m=6 \\
    &m^2_{44}=\frac{1}{12}(2\vert E\vert^2+4\Lambda)=0 ~, ~~~~~~~~~~~~ m=1 \\
    &m^2_{55}=\frac{1}{12}(6\vert E\vert^2+4\Lambda) =\frac{1}{3}\vert E\vert^2~, ~~~~~~~ m=1
\end{align}
where $m$ is the multiplicity of the eigenvalues.
There are six negative eigenvalues in Eq.(\ref{tachyonAdS1}) which do not satisfy the BF bound and therefore the corresponding AdS background is unstable.  The rest of the vacua have more complicated structure since the matrix $E_{ij}$ has more than one non-zero eigenvalues and extra non-diagonal terms appear in the mass matrix $\mathcal{M}^2$. For the second AdS vacuum in Eq.(\ref{ads1}), the eigenvalues are
\begin{align}
    &m^2_{11}=\frac{1}{6}(\vert E\vert^2+4\Lambda)=0 ~, ~~~~~~~~~~~~~~ m=9 \\
    &m^2_{22}=\frac{1}{2}m^2_{11}=0 ~, ~~~~~~~~~~~~~~~~~~~~~~~~ m=2 ~~~ \\
    &m^2_{33}=\frac{1}{6}(-2\vert E\vert^2+4\Lambda)=-\frac{1}{2}\vert E\vert^2~, ~~~ m=2 ~~~ \label{tachyon1}\\
    &m^2_{44}=\frac{1}{2}m^2_{33}=-\frac{1}{4}\vert E\vert^2~, ~~~~~~~~~~~~~~~~~ m=4 \label{tachyon2}\\
    &m^2_{55}=\frac{1}{6}(7\vert E\vert^2+4\Lambda) =\vert E\vert^2~, ~~~~~~~~~~ m=1 \\
    &m^2_{66}=\frac{1}{24}(6\vert E\vert^2+8\Lambda)=\frac{1}{6}\vert E\vert^2 ~, ~~~~~~ m=1 ~~~ \\
    &m^2_{77}=\frac{1}{24}(14\vert E\vert^2+8\Lambda)=\frac{1}{2}\vert E\vert^2 ~, ~~~~~ m=1 ~~~
\end{align}
where again two of the negative eigenvalues violate the BF bound leading to an unstable AdS. Similarly, for the Minkowski vacuum of Eq. (\ref{mink}),  the eigenvalues of the mass matrix are 
\begin{align}
    &m^2_{11}=0~, ~~~~~~~~~~~~~~ m=13 \\
    &m^2_{22}=-\frac{1}{4}\vert E\vert^2 ~, ~~~~~~ m=2 \\
    &m^2_{33}=\frac{1}{2}\vert E\vert^2~, ~~~~~~~~~ m=2 \\
    &m^2_{44}=\vert E\vert^2 ~. ~~~~~~~~~~~~ m=3
\end{align}
and therefore the Minkowski vacuum is unstable due to tachyonic modes. Lastly  the de Sitter vacuum of Eq. (\ref{ads}) is also unstable since the mass spectrum   is
\begin{align}
    &m^2_{11}=\frac{1}{6}(-\vert E\vert^2+4\Lambda)=0~, ~~~~~~~~~~~~~~ m=6 \label{dStachyon1}\\
    &m^2_{22}=\frac{1}{2}m^2_{11}=0 ~, ~~~~~~~~~~~~~~~~~~~~~~~~~~~ m=4 \label{dStachyon2}\\
    &m^2_{33}=\frac{1}{6}(5\vert E\vert^2+4\Lambda)=\vert E\vert^2~, ~~~~~~~~~~~~ m=6  \label{dStachyon3}\\
    &m^2_{44}=\frac{1}{2}m^2_{33}=\frac{1}{2}\vert E\vert^2~, ~~~~~~~~~~~~~~~~~~~~~~ m=3  \label{dStachyon4}\\
    &m^2_{55}=\frac{1}{12}(-3\vert E\vert^2+4\Lambda)=-\frac{1}{6}\vert E\vert^2 ~, ~~~~~ m=1 ~~~ \label{dStachyon5}
\end{align}
and contains one tachyonic mode. 

The tachyonic modes  in the spectrum can be lifted by considering non-constant holomorphic function ${\cal H}$. Indeed, when ${\cal H}$ has not trivial derivatives 
the equation for the scalar fluctuations takes the form 
\begin{align}
    \nabla_{\mu}\nabla^{\mu}\delta E_{ij}-\bigg\{\frac{2}{3}\Lambda+\frac{4}{\mathcal{H}+\overline{\mathcal{H}}}\left(\frac{\partial^2V}{\partial E^{ij}\partial E_{kl}}\delta E_{kl}+\frac{\partial^2V}{\partial E^{ij}\partial E^{kl}}\delta E^{kl}\right)\bigg \}=0~,
\end{align}
where the potential $V$ now is given by Eq. (\ref{potential}). Clearly, a suitable non-trivial holomorphic function ${\cal H}$ can shift the masses of the perturbations such that the tachyonic states of the spectrum are removed.  Some simple examples we worked out, indicate that when the contribution of the derivatives of ${\cal H}$ to the potential is appropriately positive, the masses are shifted accordingly. However, since   the exact form of ${\cal H}$ is not known, we can not say something more concrete at this point. 
We should  note that it is also expected that when matter fields are coupled to the theory, the above instabilities will be further  removed. 
%\textcolor{red}{In our attempt for a more complete treatment we consider the contribution of higher order terms which alter the value of the cosmological constant as previously %discussed and thus modify the mass eigenvalues. The small perturbation around the vacuum satisfies the equation
%
%The $SU(1,1)$ fields do not have kinetic terms and their contribution to the mass matrix depends on their values at the minima of the effective potential. For example, for pure real $E$ the de Sitter background discussed in Eq.(\ref{cc1}) turns to have ten tachyonic modes for $n=-1,-2,-4$. For pure imaginary $E$ there are nine tachyonic modes for $n=-1,-2,-4$ while at one of the solution of $n=-4$ where the cosmological constant becomes positively large the tachyonic modes increase to ten. For the solution in Eq.(\ref{sol22}) and specifically for $E^*=-iE$ we get de Sitter vacua for $n=-1,-2,-4$ with two tachyonic modes while for $n=-4$ there is an AdS solution with ten tachyonic modes which do not satisfy the BF bound. Comparing these results with Eq.(\ref{dStachyon1}) to (\ref{dStachyon5}) we see that the presence of higher order terms increase the instability and conclude that the vacuum stability totally depends on the choice of the function $\mathcal{H}$.}

\subsection{Non-diagonal $E_{ij}$}\label{blockdiagonal}
The previous results exclude all vacua with a diagonal form of the scalars $E_{ij}$. More general forms of $E_{ij}$ may also be considered but at the cost of increasing complexity.  
A relatively simple case that can be solved analytically is for   a non-diagonal  $E_{ij}$  of the  form
\begin{align}
E_{ij} \; : \;\;
\begin{pmatrix}
0 & 0 & 0 & 0 \\
0 & 0 & 0 & 0 \\
0 & 0 & E_{33} & E_{34} \\
0 & 0 & E_{43} & E_{44} 
\end{pmatrix} \;.
\end{align}
Analogously to the diagonal matrix with two non-zero eigenvalues, the gauge symmetry breaks in $SU(4)\rightarrow SU(2)$ and the effective potential takes the form
\begin{align}
    V_{eff}=-\frac{12}{A(S,\overline{S})}+\frac{36}{A(S,\overline{S})}\frac{\sum_{ij}\vert E_{ij}\vert^4+2\Big(\sum_i\vert E_{ii}\vert^2 \vert E_{34}\vert^2+h.c.)+(E_{44}^*E_{33}^*E_{34}^2+h.c.)\Big)}{\Big(\sum_{ij}\vert E_{ij}\vert^2\Big)^2}
\end{align}
Then, the critical points of the potential turns out to be 
\begin{align}
    &1) ~~~E_{33}E_{44}=E_{34}^2~ \label{blockconditions1}\\
    %~~~~~~~~~~~~~~~~~~~~~~~~~~~~~~~~~~:
%    \end{align}
\noalign{\text{where }}
   % \begin{align}
  &  V_{eff}(S,\overline{S})=\frac{24}{A(S,\overline{S})}~,~~~~ \Lambda=-\frac{1}{2}\sum_{ij}\vert E_{ij}\vert^2 \label{susyvac} 
%\end{align}\\
\\
\noalign{\text{and}} 
%\begin{align}
    &2)~~~E_{44}E_{34}^*=-E_{33}^*E_{34}, ~~~~ \vert E_{44}\vert^2=\vert E_{33}\vert^2 \label{blockconditions2}\\
%\end{align}
\noalign{\text{where }}
%\begin{align}
   &  V_{eff}(S,\overline{S})=\frac{6}{A(S,\overline{S})}~,~~~~ \Lambda=-\frac{1}{8}\sum_{ij}\vert E_{ij}\vert^2 \label{nonsusyvac}
\end{align}
%&E_{44}E_{34}^*=-E_{33}^*E_{34}~~~\&~~~ \vert E_{44}\vert^2=\vert E_{33}\vert^2~:~ 
The minimization of the  effective potential $V_{eff}$ leads to two different vacua of the same form as the $2\times 2$ diagonal case with a 
negative cosmological constant (corresponding to an anti-de Sitter vacuum). Note that the cosmological constant here differs to the diagonal case because of the non-diagonal elements contribution as we have noticed above. 
In the contrary, comparing the effective potential energy in Eqs.(\ref{susyvac}) and (\ref{nonsusyvac}) to the case of the $2\times 2$ diagonal matrix given in Eqs.(\ref{eff11}) and (\ref{eff22}), we see that they are equal.

\subsection{Partial Supersymmetry Breaking}
Let us now  examine whether supersymmetry is  preserved by  the vacua we found above. The Q and S- supersymmetry transformations are generated by the 
opposite chirality spinors   $\epsilon^i$ and $\eta^i$, respectively. 
The fermion shifts under $Q$ and $S$- supersymmetry, when only the $\phi^a$ and $E_{ij}$ fields are turned on, are the following\footnote{The full fermionic transformations can be found in \cite{Berg1}.}
\begin{align}
\delta_Q\Lambda_i &= 2\epsilon^{ab}\phi_a\slashed{D}\phi_b\epsilon_i +E_{ij}\epsilon^j \; ,\\
\delta_Q \chi^{ij}_{\;\;\;k} &= -\frac{1}{2}\epsilon^{ijlm}\slashed{D}E_{kl}\epsilon_m + D^{ij}_{\;\; kl}\epsilon^l~,  \label{eq2}\\
\delta_Q\psi_{\mu}^{\;\;i}&= 2(\partial_{\mu}\epsilon^i+\frac{1}{2}b_{\mu}\epsilon^i-\frac{1}{2}\omega_{\mu}\cdot\sigma\epsilon^i-V_{\mu~~j}^{~i}\epsilon^j) 
%- \gamma_{\mu}\eta^i
~,
\end{align}
and 
\begin{align}
\delta_S\Lambda_i &= 0\; ,\\
\delta_S \chi^{ij}_{\;\;\;k} &= - \frac{1}{2}\epsilon^{ijlm}E_{kl}\eta_m~,  \label{eq2}\\
\delta_S\psi_{\mu}^{\;\;i}&=  - \gamma_{\mu}\eta^i~, \label{psi3}
\end{align}
respectively. Clearly, $S$-supersymmetry is always broken since the gravitino shifts are non-zero for not trivial $\eta^i$. Similarly, the conditions for
unbroken $Q$-supersymmetry are
\begin{align}
&E_{ij}\epsilon^j=0 \; , \label{trans1}\\
&\frac{1}{4}\frac{\mathcal{DH}}{\mathcal{H}}E_{kt}E_{lf}\epsilon^{ijtf}\epsilon^l =0
%+\frac{1}{2}\epsilon^{ijlm}E_{kl}\eta_m=0 
\; ,\label{trans2}  \\
&\left(\partial_{\mu}-\frac{1}{2}\omega_{\mu}\cdot\sigma\right)\epsilon^i =0 .\label{trans3}
\end{align}
Then, the  supersymmetric  background are necessarily  Minkowski and  the scalars matrix $(E)_{ij}=E_{ij}$ should satisfy
\begin{align}
    \det E_{ij}=0.
\end{align}
In particular, the number of unbroken supersymmetries is the number of zero eigenvalues. We should also mention that another possibility is the fermionic shifts under $Q$-supersymmetry to be canceled by an $S$-fermionic shift. However, one can show that there are no no-trivial supersymmetry parameters $\epsilon^i$ and $\eta^i$ in the same direction (same index $i$) that would allow for anti-de Sitter supersymmetric backgrounds. Therefore, the only supersymmetric backgrounds in Weyl superconformal supergravity are Minkowski spacetimes.   

\section{Conclusions}
We have studied possible vacua of maximal $\mathcal{N}=4$ conformal supergravity which is the supersymmetric completion of conformal or Weyl gravity.   It is invariant under the full superconformal group $SU(2,2|4)$, the real form of $SL(4|4)$. Although such theories are considered to need UV completion, they may  emerge as a low-energy theory of string theory \cite{FKL2}. In particular, it has been claimed that it is not originating  from closed strings, but it is an effective open string theory, localized on D3-branes. 
We should notice that the superconformal symmetry we discuss here is a classical symmetry. The latter is broken by quantum effects since, although the theory is  power-counting renormalizable, it has non-vanishing  one-loop beta-functions \cite{FT2}.  Thus it suffer from  conformal anomaly so that (super)conformal  symmetry is broken. However, 
since conformal symmetry is a gauge symmetry here, it poses a thread and leads to inconsistencies \cite{FT4,T0}. 
%In other words, the theory we are discussing here cannot be the whole story.  
%The same conclusion can be drawn by considering the chiral gauge anomalies of the SU(4) R-symmetry [49] and recalling that all anomalies are accommodated in the same multiplet of the N = 4 superconformal symmetry.

We have studied the vacuum of this theory by turning on the scalars $E_{ij}$ in the $\overline{\bf 10}$ of $SU(4)$ and the scalars $\phi^a$ which parametrize 
$SU(1,1)/U(1)$ coset. The scalars $E_{ij}$ have Weyl weight $w=+1$ and therefore, their non-zero vev  breaks both conformal and $SU(4)$ symmetry. 
We have found that the theory admits de Sitter, anti-de Sitter and Minkowski vacua  determined by the vev of the scalars $E_{ij}$
and $\phi^a$. In addition, $S$-supersymmetry is always broken, whereas $Q$-supersymmetry is preserved only on Minkowski backgrounds.  The vacua we have found are  unstable as the fluctuations around them are tachyonic. This pathology indicates that  a UV completion is necessary which will remove the instability and  project out the ghost massive graviton state inherited in Weyl gravity.   

\vskip.2in
\noindent
{\bf Acknowledgement} We thank Fotis Farakos and Sergei Ketov for discussions. This research is co-financed by Greece and the European Union (European Social Fund- ESF) through the Operational Programme ``Human Resources Development, Education and Lifelong Learning 2014-2020'' in the context of the project ``Generalized Theories of Gravity" (MIS 5049089).
%This research is carried out and funded in the context of the project ``Generalized Theories of Gravity" (MIS 5049089) under the call for proposals ``Researchers' support with an emphasis on young researchers-2nd Cycle". The project is co-financed by Greece and the European Union (European Social Fund-ESF) by the Operational Programme Human Resources Development, Education and Lifelong Learning 2014-2020.


\begin{thebibliography}{99}
\bibitem{dF}
B.~de Wit and S.~Ferrara,
``On Higher Order Invariants in Extended Supergravity,''
Phys. Lett. B \textbf{81} (1979), 317-320.

\bibitem{FT}
E.~S.~Fradkin and A.~A.~Tseytlin,
``Conformal Supergravity,''
Phys. Rept. \textbf{119} (1985), 233-362.

\bibitem{FvP} D.Z. Freedman and A. Van Proeyen, ``Supergravity,'' Cambridge University Press, Cambridge U.K. (2012).

\bibitem{F1}
S.~Ferrara, M.~Kaku, P.~K.~Townsend and P.~van Nieuwenhuizen,
``Gauging the Graded Conformal Group with Unitary Internal Symmetries,''
Nucl. Phys. B \textbf{129} (1977), 125-134.

\bibitem{KvN0}
M.~Kaku, P.~K.~Townsend and P.~van Nieuwenhuizen,
``Superconformal Unified Field Theory,''
Phys. Rev. Lett. \textbf{39} (1977), 1109.

\bibitem{KvN}
M.~Kaku, P.~K.~Townsend and P.~van Nieuwenhuizen,
``Properties of Conformal Supergravity,''
Phys. Rev. D \textbf{17} (1978), 3179.

\bibitem{dWvHvP}
B.~de Wit, J.~W.~van Holten and A.~Van Proeyen,
``Transformation Rules of N=2 Supergravity Multiplets,''
Nucl. Phys. B \textbf{167} (1980), 186.

\bibitem{Berg1}
E.~Bergshoeff, M.~de Roo and B.~de Wit,
``Extended Conformal Supergravity,''
Nucl. Phys. B \textbf{182} (1981), 173-204

\bibitem{Tseyt}
I.~L.~Buchbinder, S.~M.~Kuzenko and A.~A.~Tseytlin,
``On low-energy effective actions in N=2, N=4 superconformal theories in four-dimensions,''
Phys. Rev. D \textbf{62} (2000), 045001; 
%doi:10.1103/PhysRevD.62.045001
[arXiv:hep-th/9911221 [hep-th]].

\bibitem{Buch}
I.~L.~Buchbinder, N.~G.~Pletnev and A.~A.~Tseytlin,
``\textquotedblleft{}Induced\textquotedblright{} N=4 conformal supergravity,''
Phys. Lett. B \textbf{717} (2012), 274-279
%doi:10.1016/j.physletb.2012.09.038
[arXiv:1209.0416 [hep-th]].



\bibitem{Ciceri}
F.~Ciceri and B.~Sahoo,
``Towards the full $N = 4$ conformal supergravity action,''
JHEP \textbf{01} (2016), 059, [arXiv:1510.04999 [hep-th]].

\bibitem{Butter1}
D.~Butter, F.~Ciceri, B.~de Wit and B.~Sahoo,
``Construction of all N=4 conformal supergravities,''
Phys. Rev. Lett. \textbf{118} (2017) no.8, 081602, [arXiv:1609.09083 [hep-th]].

\bibitem{Butter2}
D.~Butter, F.~Ciceri and B.~Sahoo,
``$N=4$ conformal supergravity: the complete actions,''
JHEP \textbf{01} (2020), 029, [arXiv:1910.11874 [hep-th]]

\bibitem{FKL1}
S.~Ferrara, A.~Kehagias and D.~L\"ust,
``Aspects of Weyl Supergravity,''
JHEP \textbf{08} (2018), 197, 
%doi:10.1007/JHEP08(2018)197
[arXiv:1806.10016 [hep-th]].

\bibitem{FKL2}
S.~Ferrara, A.~Kehagias and D.~L\"ust,
``Bimetric, Conformal Supergravity and its Superstring Embedding,''
JHEP \textbf{05} (2019), 100,
%doi:10.1007/JHEP05(2019)100
[arXiv:1810.08147 [hep-th]].



\bibitem{FKL3}
S.~Ferrara, A.~Kehagias and D.~L\"ust,
``Aspects of Conformal Supergravity,''
Contribution to  ISSP 2019, 
[arXiv:2001.04998 [hep-th]].


\bibitem{Anselmi1}
D.~Anselmi and M.~Piva,
``The Ultraviolet Behavior of Quantum Gravity,''
JHEP \textbf{05} (2018), 027
%doi:10.1007/JHEP05(2018)027
[arXiv:1803.07777 [hep-th]];
``Fakeons, Microcausality And The Classical Limit Of Quantum Gravity,''
Class. Quant. Grav. \textbf{36} (2019), 065010
%doi:10.1088/1361-6382/ab04c8
[arXiv:1809.05037 [hep-th]].

\bibitem{Donoghue}
J.~F.~Donoghue and G.~Menezes,
``Gauge Assisted Quadratic Gravity: A Framework for UV Complete Quantum Gravity,''
Phys. Rev. D \textbf{97} (2018) no.12, 126005
%doi:10.1103/PhysRevD.97.126005
[arXiv:1804.04980 [hep-th]].

\bibitem{Salvio}
A.~Salvio,
``Quadratic Gravity,''
Front. in Phys. \textbf{6} (2018), 77
%doi:10.3389/fphy.2018.00077
[arXiv:1804.09944 [hep-th]].


\bibitem{Stelle}
K.~S.~Stelle,
``Renormalization of Higher Derivative Quantum Gravity,''
Phys. Rev. D \textbf{16} (1977), 953-969.

%\cite{Maldacena:2011mk}
\bibitem{Maldacena}
J.~Maldacena,
``Einstein Gravity from Conformal Gravity,''
[arXiv:1105.5632 [hep-th]].
%267 citations counted in INSPIRE as of 11 Oct 2021



\bibitem{Alv}
L.~Alvarez-Gaume, A.~Kehagias, C.~Kounnas, D.~L\"ust and A.~Riotto,
``Aspects of Quadratic Gravity,''
Fortsch. Phys. \textbf{64} (2016) no.2-3, 176-189, 
%doi:10.1002/prop.201500100
[arXiv:1505.07657 [hep-th]].

\bibitem{BW}
N.~Berkovits and E.~Witten,
``Conformal supergravity in twistor-string theory,''
JHEP \textbf{08} (2004), 009,
%doi:10.1088/1126-6708/2004/08/009
[arXiv:hep-th/0406051 [hep-th]].

\bibitem{FT2}
E.~S.~Fradkin and A.~A.~Tseytlin,
``One Loop Beta Function in Conformal Supergravities,''
Nucl. Phys. B \textbf{203} (1982), 157-178.

%\cite{Ketov:2011rf}

%19 citations counted in INSPIRE as of 10 Oct 2021

%\cite{Farakos:2012qu}
\bibitem{FK2}
F.~Farakos and A.~Kehagias,
``Emerging Potentials in Higher-Derivative Gauged Chiral Models Coupled to N=1 Supergravity,''
JHEP \textbf{11}, 077 (2012)
%doi:10.1007/JHEP11(2012)077
[arXiv:1207.4767 [hep-th]].
%53 citations counted in INSPIRE as of 06 Oct 2021

%\cite{Koehn:2012ar}
\bibitem{KLO}
M.~Koehn, J.~L.~Lehners and B.~A.~Ovrut,
``Higher-Derivative Chiral Superfield Actions Coupled to N=1 Supergravity,''
Phys. Rev. D \textbf{86}, 085019 (2012)
%doi:10.1103/PhysRevD.86.085019
[arXiv:1207.3798 [hep-th]].
%77 citations counted in INSPIRE as of 06 Oct 2021

\bibitem{Ketov}
S.~V.~Ketov and N.~Watanabe,
``Cosmological properties of a generic  $\mathcal{R}^{2}$-supergravity,''
JCAP \textbf{03} (2011), 011
%doi:10.1088/1475-7516/2011/03/011
[arXiv:1101.0450 [hep-th]].

%\cite{Farakos:2013cqa}
\bibitem{FKR}
F.~Farakos, A.~Kehagias and A.~Riotto,
``On the Starobinsky Model of Inflation from Supergravity,''
Nucl. Phys. B \textbf{876} (2013), 187-200
%doi:10.1016/j.nuclphysb.2013.08.005
[arXiv:1307.1137 [hep-th]].
%172 citations counted in INSPIRE as of 10 Oct 2021

%\cite{FKP}
\bibitem{FKP}
S.~Ferrara, A.~Kehagias and M.~Porrati,
``Vacuum structure in a chiral $\mathcal{R}+\mathcal{R}^n$ modification of pure supergravity,''
Phys. Lett. B \textbf{727} (2013), 314-318
%doi:10.1016/j.physletb.2013.10.027
[arXiv:1310.0399 [hep-th]].
%24 citations counted in INSPIRE as of 10 Oct 2021

\bibitem{Mishra}
M.~Mishra and B.~Sahoo,
``Curvature squared action in four dimensional $N = 2$ supergravity using the dilaton Weyl multiplet,''
JHEP \textbf{04} (2021), 027
%doi:10.1007/JHEP04(2021)027
[arXiv:2012.03760 [hep-th]].
%1 citations counted in INSPIRE as of 11 Oct 2021
\bibitem{Dalianis:2014aya}
I.~Dalianis, F.~Farakos, A.~Kehagias, A.~Riotto and R.~von Unge,
``Supersymmetry Breaking and Inflation from Higher Curvature Supergravity,''
JHEP \textbf{01} (2015), 043
%doi:10.1007/JHEP01(2015)043
[arXiv:1409.8299 [hep-th]].
%27 citations counted in INSPIRE as of 11 Oct 2021

%\cite{Mishra:2020jlc}


\bibitem{deroo1}
M.~de Roo,
``Matter Coupling in N=4 Supergravity,''
Nucl. Phys. B \textbf{255} (1985), 515-531
%doi:10.1016/0550-3213(85)90151-8

\bibitem{deroo2}
M.~de Roo,
``Gauged N=4 Matter Couplings,''
Phys. Lett. B \textbf{156} (1985), 331-334


%\cite{deRoo:1986yw}
\bibitem{deRoo:1986yw}
M.~de Roo and P.~Wagemans,
``Partial Supersymmetry Breaking in $N=4$ Supergravity,''
Phys. Lett. B \textbf{177}, 352 (1986)
%doi:10.1016/0370-2693(86)90766-5
%31 citations counted in INSPIRE as of 15 Jul 2021

%\cite{Bellorin:2005zc}
\bibitem{Bellorin:2005zc}
J.~Bellorin and T.~Ortin,
``All the supersymmetric configurations of N=4, d=4 supergravity,''
Nucl. Phys. B \textbf{726}, 171-209 (2005)
%doi:10.1016/j.nuclphysb.2005.07.020
[arXiv:hep-th/0506056 [hep-th]].
%44 citations counted in INSPIRE as of 24 Jul 2021

%\cite{Louis:2014gxa}
\bibitem{Louis:2014gxa}
J.~Louis and H.~Triendl,
``Maximally supersymmetric AdS$_{4}$ vacua in N = 4 supergravity,''
JHEP \textbf{10}, 007 (2014)
%doi:10.1007/JHEP10(2014)007
[arXiv:1406.3363 [hep-th]].
%26 citations counted in INSPIRE as of 24 Jul 2021

%\cite{Louis:2016tnz}
\bibitem{Louis:2016tnz}
J.~Louis and S.~Lust,
``Classification of maximally supersymmetric backgrounds in supergravity theories,''
JHEP \textbf{02}, 085 (2017)
%doi:10.1007/JHEP02(2017)085
[arXiv:1607.08249 [hep-th]].
%9 citations counted in INSPIRE as of 09 Aug 2021




\bibitem{FT4}
E.~S.~Fradkin and A.~A.~Tseytlin,
``Conformal Anomaly in Weyl Theory and Anomaly Free Superconformal Theories,''
Phys. Lett. B \textbf{134} (1984), 187
%doi:10.1016/0370-2693(84)90668-3

\bibitem{T0}
A.~A.~Tseytlin,
``On divergences in non-minimal $N=4$ conformal supergravity,''
J. Phys. A \textbf{50} (2017) no.48, 48LT01
%doi:10.1088/1751-8121/aa920d
[arXiv:1708.08727 [hep-th]].


%\bibitem{Aharony:1999ti}
%O.~Aharony, S.~S.~Gubser, J.~M.~Maldacena, H.~Ooguri and Y.~Oz,
%``Large N field theories, string theory and gravity,''
%Phys. Rept. \textbf{323}, 183-386 (2000)
%doi:10.1016/S0370-1573(99)00083-6
%[arXiv:hep-th/9905111 [hep-th]].
%5073 citations counted in INSPIRE as of 08 Sep 2021


%\cite{Breitenlohner:1982bm}
\bibitem{BF1}
P.~Breitenlohner and D.~Z.~Freedman,
``Positive Energy in anti-De Sitter Backgrounds and Gauged Extended Supergravity,''
Phys. Lett. B \textbf{115} (1982), 197-201
%doi:10.1016/0370-2693(82)90643-8
%1077 citations counted in INSPIRE as of 26 Sep 2021

%\cite{Breitenlohner:1982jf}
\bibitem{BF2}
P.~Breitenlohner and D.~Z.~Freedman,
``Stability in Gauged Extended Supergravity,''
Annals Phys. \textbf{144} (1982), 249
%doi:10.1016/0003-4916(82)90116-6
%1523 citations counted in INSPIRE as of 26 Sep 2021

%(* To add *)
\begin{comment}


%\cite{Farakos:2012qu}
\bibitem{Farakos:2012qu}
F.~Farakos and A.~Kehagias,
``Emerging Potentials in Higher-Derivative Gauged Chiral Models Coupled to N=1 Supergravity,''
JHEP \textbf{11}, 077 (2012)
%doi:10.1007/JHEP11(2012)077
[arXiv:1207.4767 [hep-th]].
%53 citations counted in INSPIRE as of 24 Aug 2021

%\cite{Ooguri:2016pdq}
\bibitem{Ooguri:2016pdq}
H.~Ooguri and C.~Vafa,
``Non-supersymmetric AdS and the Swampland,''
Adv. Theor. Math. Phys. \textbf{21}, 1787-1801 (2017)
%doi:10.4310/ATMP.2017.v21.n7.a8
[arXiv:1610.01533 [hep-th]].
%228 citations counted in INSPIRE as of 24 Aug 2021

%\cite{Montero:2021otb}
\bibitem{Montero:2021otb}
M.~Montero, C.~Vafa, T.~Van Riet and G.~Venken,
``The FL bound and its phenomenological implications,''
[arXiv:2106.07650 [hep-th]].
%3 citations counted in INSPIRE as of 24 Aug 2021


\end{comment}


\end{thebibliography}
\end{document}